\documentclass[twocolumn,pra,aps,a4,floatfix,amsmath,amssymb]{revtex4}
\usepackage{amssymb}
\usepackage{epsfig}
\usepackage{hyperref}
\usepackage{color}
\usepackage{graphicx}
\usepackage{bbold}
\usepackage{amsmath}
\usepackage{amsfonts}
\usepackage{multirow}
\RequirePackage[normalem]{ulem} 
\RequirePackage{color}\definecolor{RED}{rgb}{1,0,0}\definecolor{BLUE}{rgb}{0,0,1} 
\providecommand{\DIFdeltex}[1]{{\protect\color{red}\sout{#1}}}                      
\providecommand{\DIFaddbegin}{} 
\providecommand{\DIFaddend}{} 
\providecommand{\DIFdelbegin}{} 
\providecommand{\DIFdelend}{} 
\providecommand{\DIFaddbeginFL}{} 
\providecommand{\DIFaddendFL}{} 
\providecommand{\DIFdelbeginFL}{} 
\providecommand{\DIFdelendFL}{} 
\providecommand{\DIFdel}[1]{\texorpdfstring{\DIFdeltex{#1}}{}} 
\newcommand{\DIFscaledelfig}{0.5}
\RequirePackage{settobox} 
\RequirePackage{letltxmacro} 
\newsavebox{\DIFdelgraphicsbox} 
\newlength{\DIFdelgraphicswidth} 
\newlength{\DIFdelgraphicsheight} 
\LetLtxMacro{\DIFOincludegraphics}{\includegraphics} 
\newcommand{\DIFaddincludegraphics}[2][]{{\color{blue}\fbox{\DIFOincludegraphics[#1]{#2}}}} 
\newcommand{\DIFdelincludegraphics}[2][]{
\sbox{\DIFdelgraphicsbox}{\DIFOincludegraphics[#1]{#2}}
\settoboxwidth{\DIFdelgraphicswidth}{\DIFdelgraphicsbox} 
\settoboxtotalheight{\DIFdelgraphicsheight}{\DIFdelgraphicsbox} 
\scalebox{\DIFscaledelfig}{
\parbox[b]{\DIFdelgraphicswidth}{\usebox{\DIFdelgraphicsbox}\\[-\baselineskip] \rule{\DIFdelgraphicswidth}{0em}}\llap{\resizebox{\DIFdelgraphicswidth}{\DIFdelgraphicsheight}{
\setlength{\unitlength}{\DIFdelgraphicswidth}
\begin{picture}(1,1)
\thicklines\linethickness{2pt} 
{\color[rgb]{1,0,0}\put(0,0){\framebox(1,1){}}}
{\color[rgb]{1,0,0}\put(0,0){\line( 1,1){1}}}
{\color[rgb]{1,0,0}\put(0,1){\line(1,-1){1}}}
\end{picture}
}\hspace*{3pt}}} 
} 
\LetLtxMacro{\DIFOaddbegin}{\DIFaddbegin} 
\LetLtxMacro{\DIFOaddend}{\DIFaddend} 
\LetLtxMacro{\DIFOdelbegin}{\DIFdelbegin} 
\LetLtxMacro{\DIFOdelend}{\DIFdelend} 
\DeclareRobustCommand{\DIFaddbegin}{\DIFOaddbegin \let\includegraphics\DIFaddincludegraphics} 
\DeclareRobustCommand{\DIFaddend}{\DIFOaddend \let\includegraphics\DIFOincludegraphics} 
\DeclareRobustCommand{\DIFdelbegin}{\DIFOdelbegin \let\includegraphics\DIFdelincludegraphics} 
\DeclareRobustCommand{\DIFdelend}{\DIFOaddend \let\includegraphics\DIFOincludegraphics} 
\LetLtxMacro{\DIFOaddbeginFL}{\DIFaddbeginFL} 
\LetLtxMacro{\DIFOaddendFL}{\DIFaddendFL} 
\LetLtxMacro{\DIFOdelbeginFL}{\DIFdelbeginFL} 
\LetLtxMacro{\DIFOdelendFL}{\DIFdelendFL} 
\DeclareRobustCommand{\DIFaddbeginFL}{\DIFOaddbeginFL \let\includegraphics\DIFaddincludegraphics} 
\DeclareRobustCommand{\DIFaddendFL}{\DIFOaddendFL \let\includegraphics\DIFOincludegraphics} 
\DeclareRobustCommand{\DIFdelbeginFL}{\DIFOdelbeginFL \let\includegraphics\DIFdelincludegraphics} 
\DeclareRobustCommand{\DIFdelendFL}{\DIFOaddendFL \let\includegraphics\DIFOincludegraphics} 
\RequirePackage{listings} 
\RequirePackage{color} 
\lstdefinelanguage{DIFcode}{ 
  moredelim=[il][\color{red}\sout]{\%DIF\ <\ }, 
  moredelim=[il][\color{blue}\uwave]{\%DIF\ >\ } 
} 
\lstdefinestyle{DIFverbatimstyle}{ 
	language=DIFcode, 
	basicstyle=\ttfamily, 
	columns=fullflexible, 
	keepspaces=true 
} 
\lstnewenvironment{DIFverbatim}{\lstset{style=DIFverbatimstyle}}{} 
\lstnewenvironment{DIFverbatim*}{\lstset{style=DIFverbatimstyle,showspaces=true}}{} 

\begin{document}


\title{Multiplicative Noise in Nanolaser Amplification: From Discrete Emission Events to the Additive-Noise Limit}

\author{$^{1,2}$G. Sonnino, $^1$Ph. Peeters, $^1$P. Nardone}
\affiliation{
$^1$Universit\'e Libre de Bruxelles\\
$ ^2$International SOLVAY Institutes for Physics and Chemistry\\
Bvd du Triomphe, Campus de la Plaine 224, Build. NO, 1050 Brussels, Belgium
}
\author{$^3$G.L. Lippi}
\affiliation{$^3$Universit\'e C\^ote d'Azur, UMR 7710 CNRS, Institut de Physique de Nice \\
17 rue Julien Laupr\^etre, 06200 Nice, France}
\email{\\giorgio.sonnino@ulb.be, peeters.philippe@gmail.com, \\ pasquale.nardone@ulb.be, gian-luca.lippi@univ-cotedazur.fr}


\begin{abstract}
Laser amplification kicks off with just a few photons in the cavity, a regime where standard additive Langevin noise cannot capture the real fluctuations. To understand how this amplification begins, we mapped out a single-mode photon-carrier model that links macroscopic lasers directly to high-\(\beta \) nanolasers. Stationary rate equations reveal that stimulated emission actually kicks in below the traditional gain–loss threshold. They also point to two very different crossover behaviors. In macroscopic lasers, the shift from spontaneous to stimulated emission happens across a tight pump window. In nanolasers, however, the two processes overlap across a much wider range. Because steady-state averages miss the step-by-step elementary events that build up photons, we framed the stochastic dynamics around the actual discrete transitions, deriving a state-dependent diffusion approximation. The fact that a photon is created exactly as an excited carrier drops out creates a negative photon–carrier cross-correlation. This leads directly to a correlated, multiplicative Gaussian-noise framework. We tracked these exact discrete dynamics using an event-based Laser Markov Chain. This model directly samples pump, relaxation, emission, and cavity-loss events, keeping the population counts as non-negative integers. Comparing the two shows that the multiplicative model matches the statistical behavior of photon and carrier populations perfectly, as long as a continuous diffusion description holds up. Once the emission drops into the few-photon regime, though, the discrete model becomes absolutely necessary. Eventually, at high photon counts, fluctuations shrink relative to the steady-state populations. At this point, the state-dependence of the diffusion matrix can be ignored to leading order, returning us to the classic additive Langevin description as the large-population limit of the system's stochastic dynamics.

\noindent PACS numbers: 42.55.Ah, 42.60.Mi, 42.55.Sa, 42.50.L

\end{abstract}

\maketitle

\section{Introduction}
Laser fields inevitably grow out of spontaneous emission. In large devices, the shift from weak background emission to full lasing happens so abruptly that a simple stationary threshold, paired with a standard above-threshold fluctuation model, covers most practical needs. Small cavities, however, force us to look closer at these early stages. Because a much larger fraction $\beta$ of spontaneous emission feeds directly into the lasing mode, a defining characteristic of surface-emitting architectures~\cite{Soda1979, Iga2000} among others, the resulting output curve often lacks the distinct "kink" traditionally used to pinpoint the threshold \cite{Yokoyama1989, Bjork1991, Khurgin2021, DeMartini1988, DeMartini1992, Bjork1994, Rice1994, Samuel2009, Ning2013, Deng2021, Lippi2022, Saldutti2024}. Consequently, understanding exactly what happens while the cavity holds only a few photons becomes vital for accurately describing the amplification process. Various frameworks have targeted different facets of this threshold puzzle. Statistical physics treatments focus on collective behaviour, often mapping the threshold to a phase transition in the large-system limit \cite{Haken1964, Jordan1964, Haken1964prl, Degiorgio1970, Graham1970, Grossmann1971, Dohm1972, Haken1966, Erneux1989, Tredicce2004, Marconi2024}. On the dynamics side, Maxwell-Bloch equations track the direct interaction between the field and the emitters, while the rate equations derived from them offer a streamlined way to track mean photon and carrier populations \cite{Arecchi1965mb, Narducci1988, Haken1983, Siegman1986}. Factoring spontaneous emission into the lasing mode extends these rate equations to handle smaller devices. Yet, none of these deterministic equations can independently unravel the specific sequence and timing of the individual emission events that kick off the photon buildup. Stochastic frameworks preserve far more of this granular information. Quantum models and photon-counting studies have extensively explored threshold statistics across diverse device scales \cite{Scully1966, Scully1967, Haken1970, Shore1993, Lax1967, Lax1969, Giacomotti2023, Fleck1966, Mandel1976, Stenholm1973, Weidlich1967, Gorbachev1993, Haken1967, Yamamoto1986, vanDruten2000, Woerdman2001, Kreinberg2017, Gies2007, Chow2014, Leymann2015, DAlessandro2025, Carroll2021, Wang2018, Carroll2021b, Papoff2025, Baklanov1969, Risken1966}. More recently, an event-based Laser Markov Chain (LMC) has been introduced to track individual transitions, showing strong alignment with the traditional master equation \cite{Puccioni2015, Andre2020, Bundgaard2023, Bundgaard2026} and successfully replicating the transient behaviour observed near the threshold of small lasers \cite{Wang2020}-\cite{Lippi2018}. In contrast, additive Langevin terms routinely break down near and below threshold, even if they yield perfectly reliable small-signal results at higher photon counts. Leaving aside the separate issue of pump noise \cite{Fox1984b, Cao1992, Grassi1994, Lippi1997}, our focus here is strictly on the intrinsic fluctuations generated by emission, loss, and carrier transitions within the laser itself. This distinction starts at the level of event rates. Stimulated emission scales with both the photon number $n$ and the carrier population $N$, while cavity loss depends solely on $n$. As these populations fluctuate, the transition rates and their corresponding noise strengths shift in tandem. Furthermore, a single emission event alters $n$ and $N$ simultaneously but with opposite signs. These physics demand a state-dependent, correlated noise description. However, they do not guarantee that a continuous Gaussian framework will remain valid when a single discrete event changes the total photon population by a massive percentage. To map this out, we first employ stationary rate equations to pinpoint where spontaneous and stimulated emission actively compete, tracking how this pump window scales with $\beta$. Next, we derive the explicit drift and diffusion terms directly from discrete transitions to evaluate a correlated, multiplicative Gaussian model. Finally, we establish the exact conditions under which linearising around a well-populated stationary state recovers the familiar additive Langevin equations. This progression separates three core questions that a basic average output curve cannot resolve: exactly when amplification takes off, which stochastic model holds up during its onset, and why an additive noise description becomes justified later on.

\noindent This paper is structured as follows. We begin in Section~\ref{bkgd} by establishing the background required to analyze how spontaneous and stimulated emission contribute to the average field below threshold. Section~\ref{REs} then offers a brief overview of the deterministic basic laser model, valid for a quantum-well system (QWS), alongside the coupled rate equations that govern the carrier and photon populations. From there, Section~\ref{sp-st-averages} uses the stationary solutions of these equations to explore the onset of stimulated amplification, focusing specifically on how it shifts with the spontaneous-emission factor $\beta$. We introduce the stochastic framework for the QWS model in Section~\ref{ModelNoise}, building the description directly from the fundamental discrete transitions of both populations. Within this framework, Subsections~\ref{subsec:master-to-langevin},\ref{subsec:mgwn-nanolaser} derive the multiplicative Gaussian white-noise model and detail the resulting photon-carrier anticorrelation. Conversely, Subsection~\ref{AddNoise} demonstrates how this system collapses back into Henry’s classic additive-noise description in the limit of high photon numbers. Section~\ref{Comp} covers the numerical execution of the Laser Markov Chain, which explicitly samples individual pump, relaxation, emission, and loss events, and benchmarks its predictions against the multiplicative stochastic model. Lastly, Section~\ref{concl} recaps our main findings, outlining the exact physical implications and boundaries of validity for the discrete, multiplicative, and additive stochastic regimes.


\section{Background}\label{bkgd}

The emergence of a laser field centres on a classic question: how does emission with no preferred phase spontaneously give rise to a well-defined one? Below threshold, the phase wanders completely at random over \([0,2\pi)\). Once above threshold, the field locks into a specific phase, even though residual spontaneous emission continuously coaxes it into a slow drift~\cite{Jordan1964,Haken1964prl}. In large-scale lasers, theories governing second-order phase transitions successfully map this transformation by treating the field as an order parameter controlled by the pump~\cite{Degiorgio1970,Graham1970,Grossmann1971,Dohm1972}. These frameworks predict classic hallmarks like critical slowing down and massive threshold fluctuations~\cite{Haken1966,Grossmann1971,Erneux1989,Tredicce2004,Marconi2024}. Crucially, however, such razor-sharp transitions are artefacts of the infinite-system limit. Because small lasers contain only a finite count of photons and emitters, this thermodynamic analogy cannot be adopted without heavy qualification~\cite{Rice1994}, reflecting a broader multiscale characterisation challenge inherent to emergent properties in bounded systems~\cite{Nocerino2025}. Similarly, systems containing dye media or saturable absorbers require their own dedicated frameworks, as their transitions can behave as first-order instead~\cite{Roy1979,Lugiato1978}. Maxwell-Bloch equations look at the threshold from a different angle. By coupling a classical electromagnetic field directly to quantum emitters, they view the threshold as a fundamental shift in the stability of the underlying field solutions~\cite{Arecchi1965mb,Narducci1988,Haken1983}. Adiabatically eliminating the polarisation variables yields standard rate equations for field intensity and carrier populations~\cite{Narducci1988, Haken1983, Siegman1986}. Modifying these rate equations to factor in the average spontaneous emission entering the lasing mode makes them highly effective for describing the mean output of small cavities across wide pump ranges~\cite{Yokoyama1989,Bjork1991}. Still, because they track mean populations, they remain silent on the specific, individual emission and loss events that actually build the field. To bridge this gap, standard treatments add Langevin terms to introduce noise. Most conventional calculations solve for these noise coefficients under restrictive assumptions, often evaluating them strictly at a single stationary state~\cite{Henry1983,Henry1996}. While this shortcut works remarkably well above threshold, where fluctuations are a drop in the bucket compared to the average populations, it stumbles during the initial stages of amplification. When a cavity holds very few photons, a single emission event changes the population drastically, meaning a fixed noise amplitude cannot keep up with rapidly shifting transition rates. Furthermore, continuous Gaussian trajectories can unphysically dip below zero photons. These mathematical blind spots become most glaring near and below threshold. Fully quantum models avoid these issues by quantizing the field and explicitly preserving spontaneous emission~\cite{Scully1966,Scully1967,Haken1970,Shore1993}. Pioneering studies leveraged density matrices, photon statistics, and linewidth analyses to unpack how a system approaches full lasing~\cite{Lax1967,Lax1969,Giacomotti2023,Fleck1966,Haken1964}. Related efforts expanded these boundaries to tackle multi-mode dynamics, complex photon-number distributions, and master equations~\cite{Mandel1976,Stenholm1973,Weidlich1967,Gorbachev1993}, alongside quantum Langevin approaches to isolate fluctuations~\cite{Henry1996,Haken1967,Yamamoto1986}. While these quantum methods address the blind spots of deterministic rate equations, their exact computational cost and underlying assumptions vary wildly depending on how many emitters are involved and the level of structural detail retained. The rise of nanolasers has made choosing the right framework far more urgent~\cite{vanDruten2000,Woerdman2001}. When a device relies on only a handful of active emitters, a proper model must capture the subtle correlations between emitter polarisation and the spontaneous field, correlations that simpler mean-field equations aggressively discard~\cite{Gies2007,Kreinberg2017}. Calculations of this granularity have reshaped our understanding of nanolaser photon statistics, collective emission, the murky boundary between true lasing and LED output, and the impacts of inhomogeneous broadening~\cite{Chow2014,Leymann2015,Kreinberg2017,DAlessandro2025}. Cross-scale device studies have also spotlighted unique shifts in threshold behavior and photon-number squeezing~\cite{Carroll2021,Wang2018,Carroll2021b}. For example, accounting for two-particle correlations can yield threshold states that fundamentally diverge from classic Glauber coherent states~\cite{Papoff2025}. While these developments refine how we describe the field near threshold, they do not circumvent the core necessity of analyzing the very first emission events that drive amplification. Photon-number statistics provide an excellent benchmark for vetting these competing threshold models. Researchers have long analyzed full probability distributions via Fokker-Planck equations, tracking their statistical moments, particularly the second moment, in both theory and laboratory experiments~\cite{Baklanov1969,Risken1966,Stenholm1973,Arecchi1971,Armstrong1964,Arecchi1966,Meltzer1970b,Chopra1972,Arecchi1965,Freed1965,Johnson1966,Smith1966,Davidson1967,Meltzer1970,Lax1973}. Fitting models to these empirical data points helped establish the mathematical criteria for lasing and define the quantum threshold condition itself~\cite{Sargent1970,Armstrong1965,Armstrong1965b,Haug1967,Yamamoto1992,Bjork1994,Risken1996}. Yet, while a measured distribution offers insights that an average output curve completely misses, it still represents a static snapshot; it cannot explicitly unpack the chronological order of the transitions that generated it.  Event-based simulations tackle this temporal puzzle head-on. The earliest stochastic laser models relied on discrete stimulated-emission counters and rudimentary numerical sampling schemes~\cite{Einstein1917,Rice1994,Roy-Choudhury2009,Lebreton2013,Puccioni2015}. Modern iterations have swapped these out for continuous-time transition-sampling methods~\cite{Gillespie1976,Gillespie1977,Andre2020}. Initial tests validated this approach against master-equation calculations for a single emitter before researchers extended the Laser Markov Chain framework to accommodate arbitrary emitter counts~
\cite{Bundgaard2023,Bundgaard2026}. Because this method explicitly tracks individual emission and loss events while enforcing integer photon populations, it is uniquely suited for the messy middle ground between the few-emitter regime and the macroscopic limit, a space where neither a tiny quantum toy-model nor an above-threshold Langevin calculation can tell the whole story. Experiments on micro- and nanolasers have uncovered real-world transient behaviours near threshold that event-based simulations replicate with remarkable accuracy~\cite{Wang2015,Wang2020,Wang2020pra}. One prominent interpretation views the raw measured signal as a dynamically shifting statistical blend of coherent and incoherent light during a threshold sweep~\cite{Lippi2021,Lippi2022}. Cross-method benchmarks also prove that traditional additive-noise rate equations produce fundamentally flawed statistics near and below threshold~\cite{Moerk2018,Lippi2018,Bundgaard2026}. This discrepancy does not invalidate their use for tracking minor excursions around a well-populated steady state, but it does demand a rigorous explanation of where that approximation originates and exactly where it stops matching the true microscopic physics.Transient experiments bring these physical boundaries into sharp focus. Studies of first-passage times and direct face-offs between Class A and Class B lasers demonstrate that internal relaxation scales profoundly shape threshold statistics~\cite{Roy1985,Zhu1986,Young1987,Arecchi1984,Tredicce1985,Short1982,dePasquale1986,Mecozzi1988,Arecchi1989,Meucci1992}. For instance, sweeping the pump too slowly can trigger massive, delayed turn-on responses near threshold~\cite{Scharpf1987,Arecchi1989oc,Ciofini1990,Balestri1991,Erneux1989,Erneux1991,Tredicce2004,Marconi2024}, while pump jitter warps the statistics even further~\cite{Fox1984,Fox1984b,Cao1992,Grassi1994,Lippi1997}. However, such externally driven variation is fundamentally different from the intrinsic fluctuations generated by photon emission, cavity decay, and carrier recombination. Our focus rests entirely on the latter. Because their microscopic rates depend directly on the instantaneous populations of photons and carriers, they serve as the natural launchpad for the state-dependent noise framework we build in Section~\ref{multnoise}.

\section{Deterministic model}\label{REs}

For ease of reading and reference, we now briefly recall the main features of the well-known Rate Equations model (RE) with the addition of the average contribution of the spontaneous emission~\cite{Yokoyama1989,Bjork1991}, on which the calculations of this paper are based:
\begin{eqnarray}
\label{fre1}
\dot{n} & =  & -\Gamma_c n + \beta \gamma N (n+1) \\
\label{fre2}
\dot{N} & = & R - \beta \gamma N n - \gamma N
\end{eqnarray} 
where $n$ and $N$ represent the photon and carrier number (or population inversion), respectively; $\Gamma_c$ and $\gamma$ are the relaxation rates for the intracavity photons and for the population inversion, respectively; and $R$ is the pump rate.  The fraction of spontaneous emission coupled into the lasing mode is represented by the coefficient $\beta$ ($0 \le \beta \le 1$), where the lower bound matches the thermodynamic limit, and the upper one corresponds to a device where only one electromagnetic cavity mode exists:  the lasing mode (e.g., a simple nanocavity of half-wavelength size in the emitting direction, and smaller in the other ones, if we neglect the Purcell effect~\cite{Khurgin2021}). For simplicity and without loss of generality, we do not consider here the contribution of non-radiative relaxation, unlike~\cite{Bjork1991}. Notice that while eqs.~(\ref{fre1},\ref{fre2}) are correct for any $\beta \ne 0$, the limit $\beta \rightarrow 0$ makes the coupling $\beta \gamma N n$ vanish, preventing a description of the true {\it infinite size} device with this form of the model (see~\cite{Lippi2022} for details and for the convergence between the two approaches using a different explicit form of the equations). We remark that eqs.~(\ref{fre1},\ref{fre2}) mimic the behaviour of a 4-level system and are suitable for Quantum Well devices; a similar form can be written for Quantum Dots~\cite{Moerk2018}, which are formally closer to a 3-level system, not explicitly considered here, without loss of generality. The steady-state solutions for this model read
\begin{eqnarray}
\label{nss}
\langle n \rangle & = &  \left\{ \left(\frac{r-1}{2}\right) + \sqrt{ \left(\frac{r-1}{2}\right)^2 + \beta r} \right\} \beta^{-1} \\
\label{Nss}
\langle N \rangle & = & \frac{\Gamma_c}{\beta \gamma} \frac{r}{1 + \beta \langle n \rangle} \\
\label{normthr}
r & = & \frac{R}{R_{th}}\, , \quad R_{th} = \frac{\Gamma_c}{\beta}
\end{eqnarray}
where $r = 1$ is formally considered to be the laser threshold~\cite{Rice1994}.  The normalization to the physical threshold pump rate $R_{th}$ enables comparison among different values of $\beta$, with the restriction that very small $\beta$ values \DIFdelbegin \DIFdel{to }\DIFdelend approach the thermodynamic limit ($\beta = 0$).  We stress that only the positive sign in front of the square root for the $\langle n \rangle$ solution is considered; the negative one corresponds to an unphysical solution where the average photon number is negative (see~\cite{Lippi2022} for details). The ''threshold'' definition adopted here, $R_{th}$, corresponds to the balance between gain and losses, which attributes to the (average) photon number ``at threshold'' its sole spontaneous emission contribution, as can be easily seen by taking $r = 1$ in eqs.~(\ref{nss},\ref{Nss}): 
\begin{eqnarray}
\label{thss}
\langle n \rangle_{th} = \frac{1}{\sqrt{\beta}} & \quad &
\langle N \rangle_{th} = \frac{\Gamma_c}{\gamma_{\parallel} \beta} \frac{1}{1+ \sqrt{\beta}}
\end{eqnarray}
It is useful to remark that $\langle n \rangle_{th}$ is nothing but the geometric average between the expected single photon per mode at the threshold point (QT) and the above-threshold photon number $\beta^{-1}$. 

\section{Average onset of stimulated amplification}\label{sp-st-averages}

\begin{figure}
 \centering
  \includegraphics[width=\linewidth]{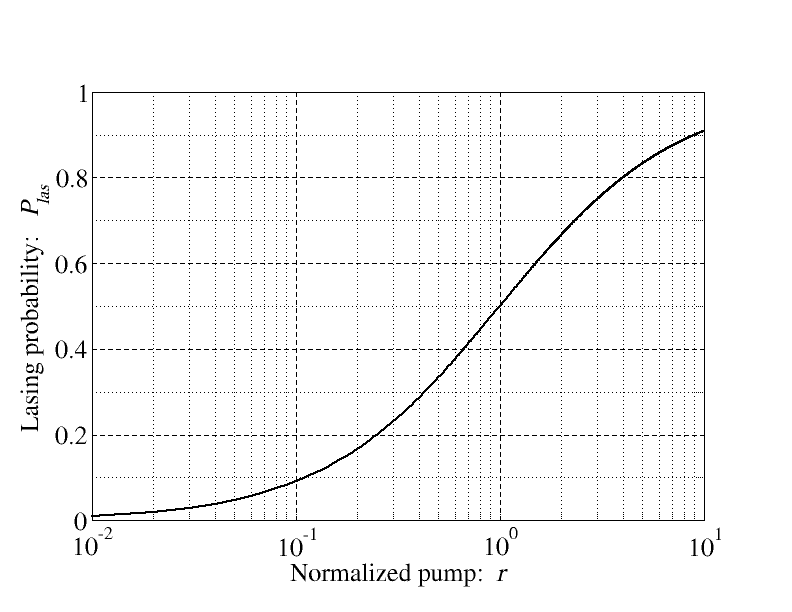} 
  \caption{Probability of initiating the lasing process as a function of the ratio $r$ between gain and losses.  $r=1$ matches the conceptual definition of laser threshold.}
  \label{fig:plas}
\end{figure}

\begin{figure}
  \centering
  \includegraphics[width=\linewidth]{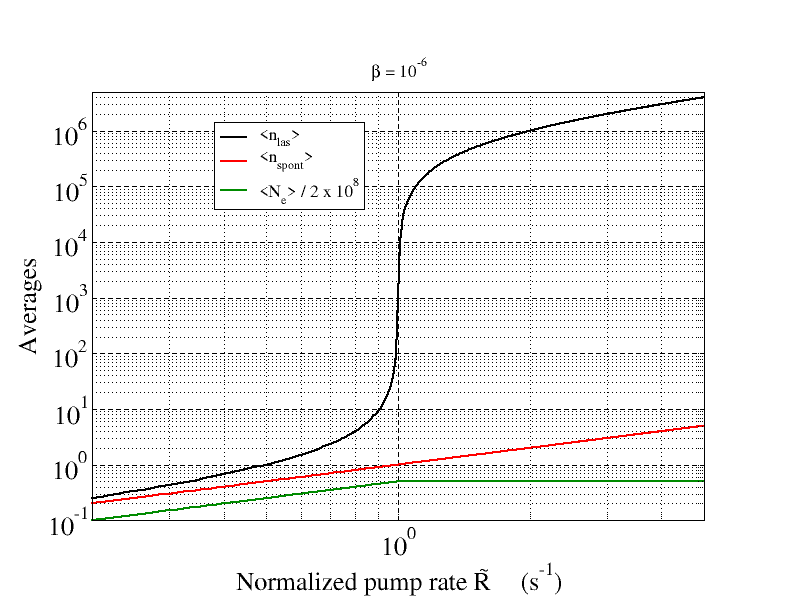} 
  \caption{Macroscopic laser:  $\beta = 10^{-6}$.  Average photon number (black) and (rescaled) emitter number (green) as a function of pump rate, normalized to threshold.  The red curve displays the sole contribution of the spontaneous emission per cavity round trip.}
  \label{fig:b-6}
\end{figure}

As discussed above, conventional descriptions of the laser threshold are based mainly on stationary, ensemble-averaged quantities, such as the mean photon number and its statistical distribution at a given pump rate. These quantities identify the progressive replacement of spontaneous emission by stimulated emission, but they do not describe how an initially random photon population develops into a coherently amplified field. In particular, an ensemble average may conceal a temporal sequence in which intervals dominated by spontaneous emission alternate with episodes of stimulated amplification \cite{Lippi2021} (in particular the discussion of Fig.~\ref{fig:b-6}), showing that the ensemble picture hides the temporal evolution of the signal where an alternation between coherent and incoherent emission may emerge, leading to an average that matches the statistical superposition (see also footnote 37 in~\cite{Lippi2022}) of different "kinds'' of photons - incoherent vs. coherent. This section aims to establish what can be inferred about the initial amplification process from average quantities alone. We first quantify the competition between stimulated emission and cavity loss as a function of the normalised pump rate. We then compare this competition with the spontaneous and stimulated contributions to the stationary photon population for different values of $\beta$. This comparison serves three purposes: it identifies the pump range in which stimulated emission becomes statistically significant, shows how that range changes with laser size, and establishes the limitations of a description based only on stationary averages. To isolate the elementary competition between amplification and loss, we consider the following balance equation for the average intracavity photon number $\langle n \rangle$:
\begin{equation}
\frac{d\langle n\rangle}{dt}=-\kappa\langle n\rangle+R_{\mathrm{sp}}+G\langle n\rangle 
\end{equation}
where $\kappa$ is the total photon-loss rate, $R_{\mathrm{sp}}$ is the spontaneous-emission rate into the lasing mode, and $G$ is the stimulated gain per unit time. This reduced equation does not describe the complete threshold dynamics. It provides, however, a direct measure of the probability that a photon already present in the cavity produces a stimulated-emission event before it is lost. Equilibrium provides the average photon number
\begin{equation}
\langle n \rangle  =  \frac{R_{sp}}{\kappa - G} 
\end{equation}
The denominator identifies the threshold through the (idealised) divergence of the expression, while no particular significance is to be attributed to the sign of $\langle n \rangle$ when gain exceeds total losses, since we focus on the parameter region below threshold. A photon achieves a round trip inside the cavity, of length $L$, in a time 
\begin{equation}
\Delta t  = 2 L \frac{n_{eff}}{c} 
\end{equation}
$n_{eff}$ represents the effective medium's refractive index and $c$ the speed of light in vacuum) and experiences stimulated gain and loss with probabilities $\mathcal{P}_{stim} = G \Delta t$ and $\mathcal{P}_{loss} = \kappa \Delta t$, respectively.  The ratio between these probabilities 
\begin{eqnarray}
r & = & \frac{\mathcal{P}_{stim}}{\mathcal{P}_{loss}} = \frac{G}{\kappa}
\end{eqnarray}
measures the respective strength of the two mechanisms and sets the threshold at $r = 1$. The probability of initiating lasing is given by the probability of gain normalized to the total probability per round trip
\begin{equation}
\mathcal{P}_{las}  =  \frac{\mathcal{P}_{stim}}{\mathcal{P}_{stim} + \mathcal{P}_{loss}} =  \frac{G}{G+\kappa} = \frac{r}{r+1}
\end{equation}
Thus, at $r = 1$ (threshold) the probability of initiating lasing is 50\%. The dependence of $\mathcal{P}_{las}$ on $r$ is plotted in Fig.~\ref{fig:plas}, where we see that the probability is non-zero even before threshold, although it is below 20\% when the pump rate is set at 30\% of its threshold value.  The threshold definition, where "{\it half of the photons emitted into the mode will be emitted coherently, the other half will be added noncoherently}''~\cite{Yamamoto1992}, matches what is also called the ``{\it quantum definition} where {\it the mean photon number at threshold is unity for a truly ideal laser}''~\cite{Yamamoto1992,Bjork1994}, i.e., a laser devoid of non-radiative relaxation. "\textit{At this rate, the photon emission rate is exactly the rate at which new carriers must be injected to compensate for the loss}''~\cite{Yamamoto1992}.
We now turn to the analysis of the predictions that the rate equations give for the averages of both photon number and emitter number for different laser sizes. Fig.~\ref{fig:b-6} shows the photon (black) and the emitter (green) numbers as a function of the pump rate, normalized to threshold for a macroscopic laser ($\beta = 10^{-6}$); the emitter number curve is rescaled for graphical purposes. For clarity, in the following we will entirely neglect the background radiation whose contribution at room temperature and $\lambda = 1 \mu m$ can be estimated, from Bose-Einstein statistics, around $2 \times 10^{-21}$ photons at the frequency of interest.
Comparison between the average intracavity photon population (black) and the spontaneous contribution (red) shows that the latter constitutes the majority of the photon number at low pump. Quantitatively, at $r = 0.2$, less than 6\% of the photon number is constituted by stimulated photons; Fig.~\ref{fig:plas} shows that at the same pump rate one expects a probability of about 17\% of initiating lasing action. The stimulated component, however, increases rapidly, and at $r =0.5$ doubles the number of intracavity photons, as compared to the spontaneous component, while the probability of lasing initiation is $\frac{1}{3}$. The growth increases nonlinearly to reach a factor of $10^3$ at $r = 1$, with a 50\% probability of lasing, which matches a photon number equal to $\beta^{\frac{1}{2}}$. It is interesting to remark that the threshold value matches $\langle n_{spont} \rangle = 1$ (the quantum threshold~\cite{Yamamoto1992,Bjork1994}), commonly interpreted as the spontaneous average photon number (in each mode) for the occurrence of the lasing transition, and that, as expected, the emitter population $\langle N \rangle$ stops growing at this same pump value, since from this pump value on the energy supplied to the laser, in the form of inverted carriers, is transformed into photon number.
\begin{figure}
  \centering
  \includegraphics[width=\linewidth]{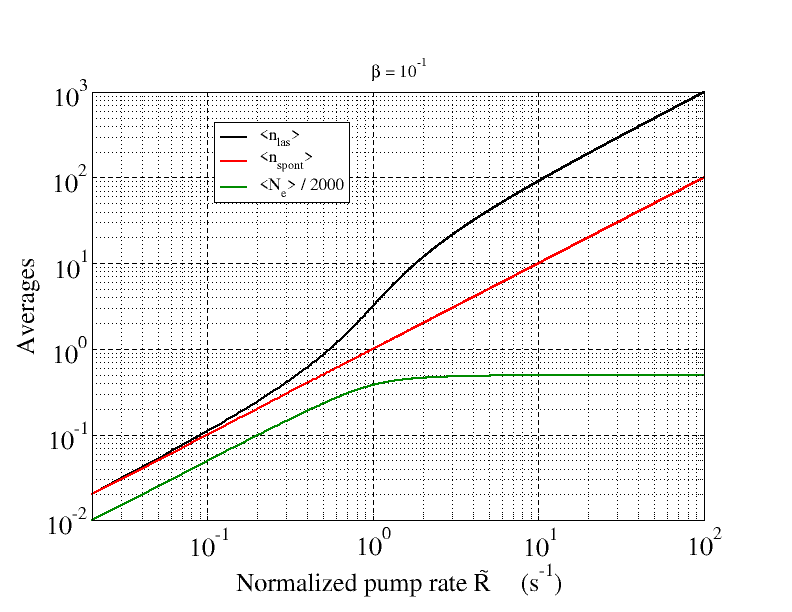} 
  \caption{Nanolaser:  $\beta = 10^{-1}$.  Average photon number (black) and (rescaled) emitter number (green) as a function of pump rate, normalized to threshold.  The red curve displays the sole contribution of the spontaneous emission per cavity round trip.}
  \label{fig:b-1}
\end{figure}

\noindent Inspection of the same curves for $\beta = 10^{-1}$ (a nanolaser) produces the well-known much smoother transition from the dominant spontaneous to stimulated emission (Fig.~\ref{fig:b-1}).  While at very low pump (e.g., $r = 0.02$) the photons are virtually all spontaneous, at $r = 0.2$ the stimulated fraction amounts to approximately 21\% (vs. $<$ 6\% for the macrolaser). This fact is easily understood in terms of available e.m. cavity modes; in the nanolaser they are in much smaller number, thus it is easier to obtain a larger average contribution of the stimulated fraction below threshold, thanks to the fewer available modes. The growth is, however, much slower: the stimulated fraction is approximately 70\% of the total at $r = 0.5$ (vs. 100\% for the macrolaser) and amounts to about a factor of 3 (more precisely $\beta^{\frac{1}{2}}$, again) at threshold. As a consequence, the near totality of the emission becomes stimulated (parallel black and red curves) only about ten times above threshold, as opposed to about twice threshold in the macrolaser. The preceding analysis identifies the average conditions under which stimulated emission begins to compete with spontaneous emission and cavity loss. It also shows that this competition starts below the nominal gain-loss threshold and extends over a pump interval whose width depends strongly on $\beta$. For a macroscopic laser, the crossover is confined to a narrow region, whereas in a nanolaser the spontaneous and stimulated contributions coexist over a much broader range. These stationary averages do not reveal the temporal ordering of the underlying events. They cannot determine whether the measured photon population results from a nearly continuous weak amplification or from intermittent sequences in which a spontaneous photon seeds a burst of stimulated emission. Experiments on mesoscale lasers, together with Laser Markov Chain simulations, indicate that the latter dynamics can play an important role near threshold \cite{Puccioni2015,Wang2015, Lippi2021}. A description of the initial buildup must therefore retain the dependence of the fluctuations on the instantaneous photon and carrier populations. This state dependence is the subject of Section~\ref{multnoise}. 

\section{Multiplicative noise}\label{multnoise}\label{ModelNoise}

The limitations of an additive-noise description near threshold can be traced to the microscopic processes that generate the fluctuations. Stimulated emission, introduced by Einstein in 1917 \cite{Einstein1917}, occurs at a rate proportional to both the available population inversion and the number of photons already present in the cavity. In the notation of Eqs.~(\ref{fre1}) and (\ref{fre2}), this rate is $\beta\gamma Nn$. The bilinear deterministic term provides a useful first indication of the state dependence, although it does not by itself determine the noise. Writing $n=\langle n\rangle+\delta n$ and $N=\langle N\rangle+\delta N$ gives, to first order,
\begin{equation}
    \delta(Nn)\simeq \langle N\rangle\,\delta n
    +\langle n\rangle\,\delta N
    \label{eq:gain-linearization}
\end{equation}
For ensemble averages, one must additionally retain the carrier-photon covariance,
\begin{equation}
    \langle Nn\rangle=\langle N\rangle\langle n\rangle
    +\operatorname{Cov}(N,n)
    \label{eq:gain-covariance}
\end{equation}
Neglecting the last term is the mean-field closure whose consequences for threshold indicators were discussed in Ref.~[93]. Equation~\eqref{eq:gain-linearization} therefore describes the linearized deterministic feedback, whereas the multiplicative character of the noise follows from the statistics of the underlying emission, decay, and injection events. Spontaneous emission into the lasing mode contributes the additional rate $\beta\gamma N$, so that the total rate of photon-producing emission events is
\begin{equation}
    a_{\mathrm{em}}(n,N)=\beta\gamma N(n+1)
    \label{eq:emission-propensity}
\end{equation}
This expression contains the physical origin of the stochastic state dependence: the frequency of the random emission events changes with the instantaneous values of $n$ and $N$.

\subsection{From the master equation to multiplicative noise}\label{subsec:master-to-langevin}\label{FPModelNoise}

The stochastic evolution of a system with discrete states can be described by a continuous-time Markov process. Let $P(n,t)$ be the probability that the system occupies state $n$ at time $t$, and let $W(n\mid n')$ denote the transition rate from state $n'$ to state $n$. Conservation of probability then gives the master equation
\begin{equation}
    \frac{dP(n,t)}{dt}\!=\!\sum_{n'}\left[W(n\!\mid\! n')P(n',t)\!-\!W(n'\!\mid\! n)P(n,t)\right]\label{eq:master-equation-noise}
\end{equation}
The first term accounts for transitions into state $n$, whereas the second accounts for transitions out of it. Equation~\eqref{eq:master-equation-noise} is exact for the chosen set of states and transition rates and retains the discrete character of the underlying events. To obtain a continuous approximation, we introduce the intensive variable $x=n/\Omega$ where $\Omega$ is a large system-size parameter, such as the number of emitters or an effective volume. A transition $n\rightarrow n'$ produces the increment $\Delta x_{n'n}=(n'-n)/\Omega$. The probability density of the continuous variable, denoted by $p(x,t)$, is related asymptotically to the discrete probability through $p(x,t)\simeq\Omega P(n,t)$. Expanding the master equation in powers of the small increments $\Delta x_{n'n}$ gives the Kramers--Moyal expansion
\begin{equation}
    \frac{\partial p(x,t)}{\partial t}
    =\sum_{k=1}^{\infty}
    \frac{(-1)^k}{k!}
    \frac{\partial^k}{\partial x^k}
    \left[M^{(k)}(x)p(x,t)\right]
    \label{eq:kramers-moyal-expansion}
\end{equation}
where
\begin{equation}
    M^{(k)}(x)
    =\sum_{n'}\left(\Delta x_{n'n}\right)^k W(n'\mid n)
    \label{eq:jump-moments}
\end{equation}
is the $k$th jump moment per unit time, evaluated at $n=\Omega x$. If individual jumps are small on the scale over which $p(x,t)$ varies and sufficiently many events occur during the coarse-graining interval, the terms of order $k\geq3$ may be neglected. Defining the drift and diffusion coefficients as
\begin{equation}
    A(x)=M^{(1)}(x),
    \qquad
    B(x)=M^{(2)}(x)
    \label{eq:drift-diffusion-definitions}
\end{equation}
one obtains the Fokker--Planck equation
\begin{equation}
    \frac{\partial p(x,t)}{\partial t}
    =-\frac{\partial}{\partial x}
        \left[A(x)p(x,t)\right]
    +\frac{1}{2}\frac{\partial^2}{\partial x^2}
        \left[B(x)p(x,t)\right]
    \label{eq:fokker-planck-general}
\end{equation}
Here $A(x)$ gives the deterministic rate of change, while $B(x)$ determines the variance accumulated per unit time. The Fokker--Planck equation~\eqref{eq:fokker-planck-general} is equivalent to the It\^o stochastic differential equation
\begin{equation}
    dx=A(x)\,dt+\sqrt{B(x)}\,dW(t)
    \label{eq:ito-langevin-general}
\end{equation}
where $dW(t)$ is a Wiener increment satisfying
\begin{equation}
    \langle dW(t)\rangle=0,
    \qquad
    \langle dW(t)^2\rangle=dt
    \label{eq:wiener-statistics}
\end{equation}
Equivalently, introducing Gaussian white noise $\xi(t)=dW/dt$, Eq.~\eqref{eq:ito-langevin-general} may be written as
\begin{equation}
    \dot{x}\!=\!A(x)\!+\!\sqrt{\!B(x)}\xi(t),
    \ 
    \langle\xi(t)\rangle\!=\!0,
    \ 
    \langle\xi(t)\xi(t')\rangle\!\!=\!\delta(t\!-\!t')
    \label{eq:white-noise-langevin-general}
\end{equation}
The noise is additive only when $B$ is independent of the state. If $B=B(x)$, the noise amplitude changes with the instantaneous state and the noise is multiplicative. This dependence is not introduced phenomenologically: it follows from the state dependence of the transition rates in the master equation. For the laser problem, the state is multidimensional because both the photon number and the carrier population fluctuate. Writing ${\bf x}=(n,N)^{\mathrm T}$ and labeling the elementary processes by $j$, with state-change vectors ${\boldsymbol\nu}_j$ and propensities $a_j({\bf x})$, the drift vector and diffusion matrix are
\begin{align}
    A_i({\bf x})
    &=\sum_j \nu_{ij}a_j({\bf x}),
    \label{eq:multivariate-drift}\\
    D_{ik}({\bf x})
    &=\sum_j \nu_{ij}\nu_{kj}a_j({\bf x})
    \label{eq:multivariate-diffusion}
\end{align}
The corresponding Fokker-Planck equation reads
\begin{equation}
    \frac{\partial p}{\partial t}
    =-\sum_i\frac{\partial}{\partial x_i}\left(A_i p\right)
    +\frac{1}{2}\sum_{i,k}
    \frac{\partial^2}{\partial x_i\partial x_k}
    \left(D_{ik}p\right)
    \label{eq:multivariate-fokker-planck}
\end{equation}
and its It\^o Langevin representation is
\begin{equation}
    d{\bf x}={\bf A}({\bf x})\,dt
    +{\bf S}({\bf x})\,d{\bf W},
    \qquad
    {\bf S}({\bf x}){\bf S}^{\mathrm T}({\bf x})
    ={\bf D}({\bf x})
    \label{eq:multivariate-langevin}
\end{equation}
The off-diagonal elements of ${\bf D}$ describe correlations generated when a single event changes more than one dynamical variable. In particular, an emission event into the lasing mode simultaneously increases $n$ and decreases $N$. The state dependence and cross correlations of the laser noise therefore follow directly from the same transition rates that determine the average rate equations. The passage from Eq.~\eqref{eq:master-equation-noise} to Eq.~\eqref{eq:multivariate-langevin} is a diffusion approximation, not an exact rewriting of the discrete dynamics. It becomes reliable when the relevant populations and event counts are large. Near and below threshold, where the intracavity photon number may be of order unity, higher jump moments and the non-negative, integer-valued character of $n$ cannot in general be neglected. In that regime, the original master equation or an equivalent event-based Markov-chain description remains the appropriate starting point.

\subsection{Multiplicative Gaussian white-noise model}\label{subsec:mgwn-nanolaser}

We consider now the case of a quantum well system where the evolution equations for photons and carriers are governed by the following nonlinear stochastic equations:
\begin{eqnarray}
\label{seqn1}
\dot{n} & =  & -\Gamma_c n + \beta \gamma N (n+1)+G_n(t) \\
\label{seqN2}
\dot{N} & = & R - \beta \gamma N n - \gamma N+G_N(t)
\end{eqnarray} 
We interpret the stochastic differential equations~(\ref{seqn1}) and (\ref{seqN2}) in the Stratonovich sense. The associated probability density $P(n, N,t)$ then satisfies the corresponding Fokker–Planck equation. The diffusion approximation introduced in the previous section leads naturally to state-dependent fluctuations. In the nanolaser rate equations, the stochastic forces acting on the photon number $n$ and the carrier population $N$ are denoted by $G_n(t)$ and $G_N(t)$, respectively. Our phenomenological \textit{Multiplicative Gaussian White-Noise} (MGWN) model assumes that the mean and the local fluctuation amplitude of each force scale with the instantaneous population on which it acts. Our model should combine Gaussian white noise with a non-zero expectation value for $G_j(t)$ (with $j=n, N$). To this end, we define $G_j(t)=j\zeta_j(t)$ where $\zeta_j(t)$ is a Gaussian noise with mean $\mu\neq 0$ and variance $\eta^2$, i.e., $\zeta_j(t)\sim N(\mu,\eta^2)$. Then, the expectation value $<G_j(t)>$ scales as $\mu_j$ and the variance of $G_j(t)$ scales as $\eta^2$. Hence, these two scalings are compatible with each other by setting $\mu=\eta$. We finally get:
\begin{align}\label{model1}
&<G_j(t)>=\eta<j>\nonumber\\
&<G_j(t)<G_j(t')>=\tau_j\eta^2\delta(t-t')\ \ {\rm with}\  j=n,N
\end{align}
where $\tau_j$ are the \textit{relaxation characteristic times}. For our quantum well system, we therefore prescribe
\begin{align}
    &\langle G_n(t)\rangle = \eta_1\langle n(t)\rangle,\quad \langle G_N(t)\rangle = \eta_2\langle N(t)\rangle, \label{eq:mgwn-means}\\
   & \langle G_n(t) G_n(t')\rangle=\tau_n\eta_1^2\langle n^2(t)\rangle\delta(t-t'),\nonumber\\
    & \langle G_N(t)G_N(t')\rangle=\tau_N\eta_2^2\langle N^2(t)\rangle\delta(t-t'),
    \label{eq:mgwn-autocorrelations}\\
    &\langle G_n(t) G_N(t')\rangle=-\eta_3^2\sqrt{\tau_n\tau_N}\,
      \langle n(t)N(t)\rangle\delta(t-t')
    \label{eq:mgwn-crosscorrelation}
\end{align}
Here $\tau_n$ is a characteristic photon time scale, which may be chosen on the order of the cavity lifetime, and $\tau_N$ is the carrier relaxation time. The parameters $\eta_1$ and $\eta_2$ control the strengths of the photon and carrier fluctuations, whereas $\eta_3$ controls their anticorrelation. These parameters are undetermined at this stage. The negative sign in Eq.~\eqref{eq:mgwn-crosscorrelation} reflects the fact that an emission event increases the photon population while depleting the carrier population. We introduce the positive scalar diffusion magnitudes
\begin{equation}
    D_n=\frac{1}{2}\eta_1^2\tau_n,
    \ 
    D_N=\frac{1}{2}\eta_2^2\tau_N,
    \ 
    D_{nN}=\frac{1}{2}\eta_3^2\sqrt{\tau_n\tau_N}
    \label{eq:mgwn-diffusion-amplitudes}
\end{equation}
with $D_{nN}^2\leq D_nD_N$. This is the diffusion form of the Cauchy-Schwarz inequality. Physically, the cross-diffusion strength cannot exceed the geometric mean of the separate photon and carrier diffusion strengths. If $D_{nN}=0$, the two fluctuations are uncorrelated. If $0<D_{nN}<\sqrt{D_nD_N}$, they are only partially anticorrelated because photon and carrier noise also contain contributions from independent processes, such as cavity loss and carrier injection. The limiting case $D_{nN}^2=D_nD_N$ corresponds to perfect anticorrelation, $\rho_{nN}=-1$. The diffusion matrix then has zero determinant and only one independent stochastic direction: after normalization by their respective fluctuation amplitudes, every photon-number increase is accompanied by a corresponding carrier-population decrease. A value $D_{nN}^{2}>D_nD_N$ would make one eigenvalue of the diffusion matrix negative, implying a negative variance for some linear combination of $n$ and $N$ and therefore an unphysical stochastic model.

\subsubsection{Construction from two independent gaussian sources}\label{2GS}

The correlations in Eqs.~\eqref{eq:mgwn-means}--\eqref{eq:mgwn-crosscorrelation} can be generated from two statistically independent Gaussian white-noise sources $\xi_n(t)$ and $\xi_N(t)$. The shifted processes $\xi_n$ and $\xi_N$ have unit mean; their centered parts are white and mutually independent. The photon and carrier forces may then be represented as
\begin{align}
    G_n(t)
    &=n(t)\left[\eta_a\xi_n(t)+\eta_b\xi_N(t)\right]
    \label{eq:mgwn-Gn-decomposition}\\
    G_N(t)
    &=N(t)\left[\eta_b\xi_n(t)+\eta_c\xi_N(t)\right]
    \label{eq:mgwn-GN-decomposition}
\end{align}
with $\xi_1$ and $\xi_2$ independent normalized white noises. The coefficient $\eta_b$ couples both dynamical variables to the same two sources and therefore permits a nonzero cross-correlation. In an It\^o implementation, the instantaneous values of $n$ and $N$ multiply Wiener increments that are independent of the state at the beginning of the time step. Matching the prescribed means gives
\begin{equation}
    \eta_a+\eta_b=\eta_1, \qquad \eta_b+\eta_c=\eta_2\label{eq:mgwn-mean-matching}
\end{equation}
Introducing the ratio of characteristic times
\begin{equation}
    \chi=\frac{\tau_N}{\tau_n}
    \label{eq:mgwn-chi}
\end{equation}
matching the two variances and the cross-covariance yields
\begin{align}
    c_1\eta_a^2+c_2\chi\eta_b^2
    &=\eta_1^2,
    \label{eq:mgwn-condition-one}\\
    c_1\eta_b^2+c_2\chi\eta_c^2
    &=\chi\eta_2^2,
    \label{eq:mgwn-condition-two}\\
    c_1\eta_a\eta_b+c_2\chi\eta_b\eta_c
    &=-\eta_3^2\sqrt{\chi}
    \label{eq:mgwn-condition-three}
\end{align}
The first two relations fix the diagonal noise strengths, whereas the third fixes the photon--carrier anticorrelation. Using $\eta_a=\eta_1-\eta_b$ and $\eta_c=\eta_2-\eta_b$, the coefficients $c_1$ and $c_2$ can be expressed in terms of $\eta_b$ as
\begin{align}
    c_1
    &=\frac{
        \eta_1^2(\eta_2-\eta_b)^2
        -\chi\eta_2^2\eta_b^2
    }{
        (\eta_1-\eta_b)^2(\eta_2-\eta_b)^2-\eta_b^4
    },
    \label{eq:mgwn-c1-solution}\\
    c_2
    &=\frac{
        \chi\eta_2^2(\eta_1-\eta_b)^2
        -\eta_1^2\eta_b^2
    }{
        \chi\left[(\eta_1-\eta_b)^2(\eta_2-\eta_b)^2-\eta_b^4\right]
    }
    \label{eq:mgwn-c2-solution}
\end{align}
The remaining coefficient $\eta_b$ is determined by
\begin{multline}
    \eta_b(\eta_1-\eta_b)
    \left[\eta_1^2(\eta_2-\eta_b)^2
          -\chi\eta_2^2\eta_b^2\right]
    \\
    +\eta_b(\eta_2-\eta_b)
    \left[\chi\eta_2^2(\eta_1-\eta_b)^2
          -\eta_1^2\eta_b^2\right]
    \\
    =-\eta_3^2\sqrt{\chi}\,
    \left[(\eta_1-\eta_b)^2(\eta_2-\eta_b)^2-\eta_b^4\right]
    \label{eq:mgwn-etab-equation}
\end{multline}
Among the algebraic roots of Eq.~\eqref{eq:mgwn-etab-equation}, only those for which the denominator in Eqs.~\eqref{eq:mgwn-c1-solution} and \eqref{eq:mgwn-c2-solution} is nonzero and $c_1,c_2\geq0$ define real Gaussian amplitudes. The positivity condition $D_{nN}^2\leq D_nD_N$ must also hold. The MGWN model is a continuous Gaussian approximation. It captures the scaling of fluctuation amplitudes with $n$ and $N$ and permits the photon-carrier anticorrelation to be imposed explicitly. It does not, however, restore the discreteness of emission and loss events. Its quantitative use near $n=0$ therefore remains subject to the same population-boundary limitations discussed above; in the few-photon regime, the event-based Markov description provides the more reliable reference.

\paragraph Discussion: Breakdown of the diffusion approximation near threshold. The multiplicative Langevin equations are themselves an approximation to the discrete jump process. Their derivation requires each active channel to generate many events over a coarse-graining interval $\Delta t$, while the fractional change of $n$ and $N$ over that interval remains small. These requirements are incompatible when the intracavity population is of order unity. A single emission or loss event then changes $n$ by an amount comparable to its current value, and the distribution of increments is strongly non-Gaussian. This observation clarifies the numerical failure reported in Ref.~\cite{Bundgaard2026}. Gaussian increments have unbounded support and can drive a continuous photon variable below zero, especially below threshold, where the physical state lies close to the boundary $n=0$. Clipping negative values, reflecting them at the boundary, or setting the noise amplitude to zero after a negative step may prevent numerical divergence, but each prescription changes the transition probabilities and hence the photon statistics. State-dependent noise is therefore physically necessary, but a multiplicative Gaussian equation does not by itself restore the discrete, non-negative dynamics in the few-photon regime. The Laser Markov Chain, implemented through exact event sampling or a controlled tau-leaping scheme, retains integer populations and the correct boundary automatically \cite{Bundgaard2026}. The same limitation holds for the carrier population when the number of active emitters is small. In a macroscopic device, however, $N$ may already be large below threshold even though $n$ is not. The validity of a diffusion model must therefore be assessed channel by channel rather than inferred from the nominal laser size alone. In particular, a large carrier reservoir does not justify a Gaussian approximation for photon loss and stimulated-emission events when only a few photons occupy the cavity.

\subsection{Recovery of additive noise at large photon number}\label{AddNoise}

\begin{figure}
  \centering
  \includegraphics[width=\linewidth]{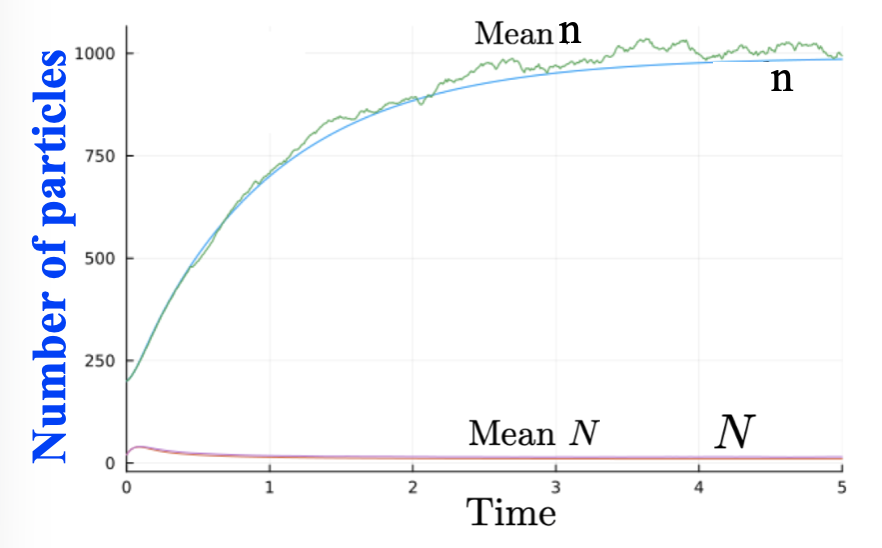} 
  \caption{Nanolaser: $\beta = 10^{-1}$, the values of the other parameters are $\Gamma_c=10^{9}sec^{-1}$, $\gamma=\gamma_c=10^9sec^{-1}$, and $R=10^{12} electron\ sec^{-1}$. This figure shows the evolution of the photon and carrier populations obtained with Henry's additive-noise model. The difference between the stochastic and deterministic carrier populations is barely visible.}
  \label{figSNAdd}
\end{figure}

\begin{figure}
  \centering
  \includegraphics[width=\linewidth]{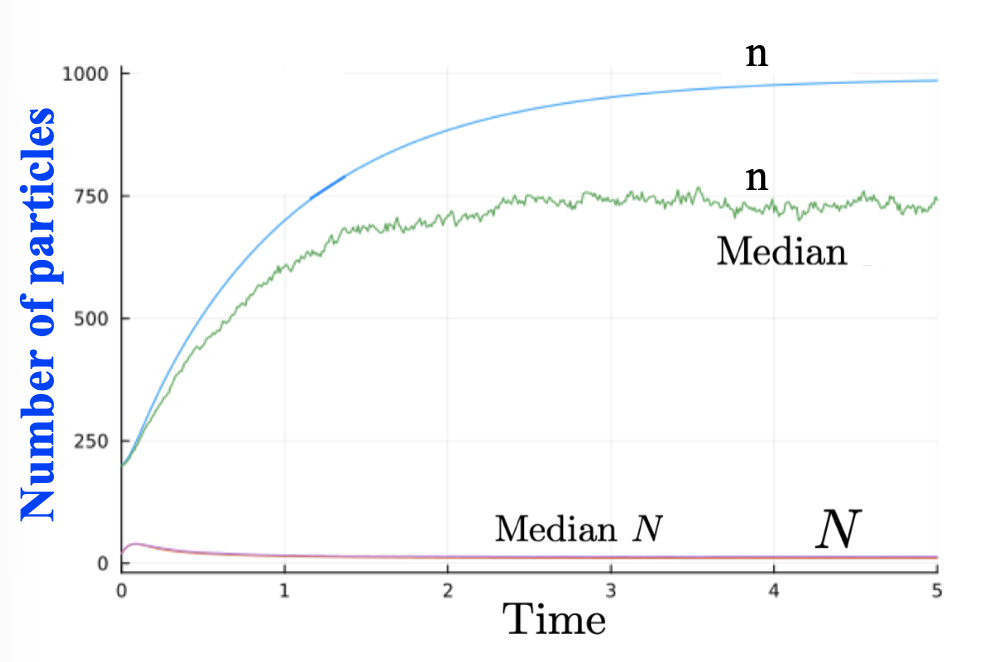} 
  \caption{Nanolaser:  $\beta = 10^{-1}$. Evolution of the photon and carrier populations obtained by our multiplicative noise model. The values of the stochastc parameters are $\eta_1=10^9sec^{-1}$,  $\eta_2=10^9sec^{-1}$,  $\eta_3=10^5sec^{-1}$, and $\chi=1.0$.}
  \label{figSNMultPh}
\end{figure}

\begin{figure}
  \centering
  \includegraphics[width=\linewidth]{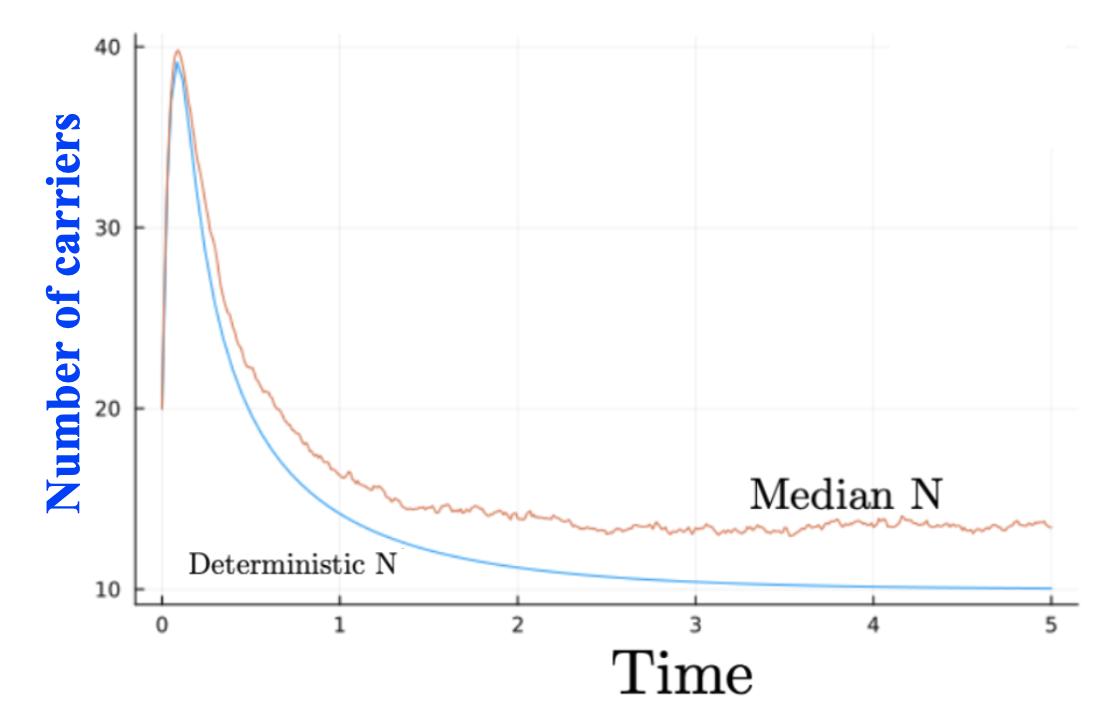} 
  \caption{Nanolaser:  $\beta = 10^{-1}$. Evolution of the carrier population obtained with our multiplicative model. The values of the stochastic parameters are the same as those mentioned in Fig.~\ref{figSNMultPh}.}
  \label{figSNMultCa}
\end{figure}

\begin{figure}
  \centering
  \includegraphics[width=\linewidth]{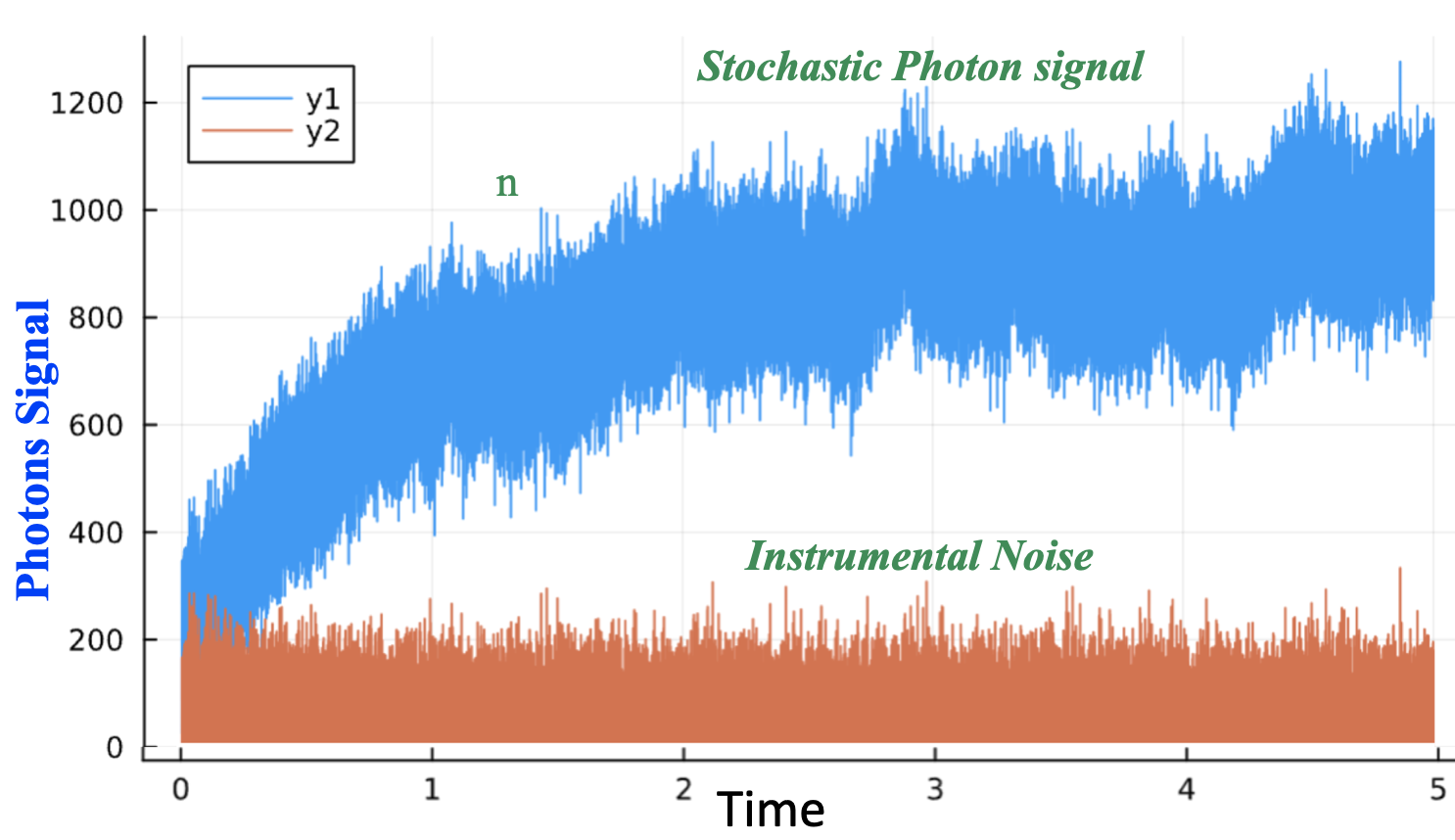} 
  \caption{Nanolaser:  $\beta = 10^{-1}$. Display of the fluctuations of the stochastic photon signal generated by the QWS model with multiplicative noise. The blue shaded region represents the statistical spread of the photon signal, whereas the brown region indicates the intrinsic instrumental noise. The values of the stochastic parameters are the same as those mentioned in Fig.~\ref{figSNMultPh}.}
  \label{figStSignals}
\end{figure}

Above threshold, the event rates become large and their relative variations over a correlation time become small. To see how additive noise emerges, we write ${\bf x}=\overline{\bf x}+\delta{\bf x}$, where $\overline{\bf x}$ is the stationary solution of Eqs.~(3) and (4), and denote the deterministic drift by ${\bf A}({\bf x})$. Linearization gives
\begin{equation}
    d\delta{\bf x}={\bf J}\,\delta{\bf x}\,dt
    +{\bf B}(\overline{\bf x})\,d{\bf W}
    +\mathcal{O}\!\left(\delta{\bf x}\,d{\bf W},\delta{\bf x}^{2}\right),
    \label{eq:linear-noise-limit}
\end{equation}
where ${\bf J}=\partial{\bf A}/\partial{\bf x}|_{\overline{\bf x}}$ and
${\bf B}{\bf B}^{\mathrm T}={\bf D}$. At leading order, the noise matrix is evaluated at the steady state and is therefore constant. The fluctuations then obey an Ornstein-Uhlenbeck process with additive Gaussian forcing. This is the linear-noise approximation used in conventional small-signal analyses. Equation~\eqref{eq:linear-noise-limit} explains both the usefulness and the limitations of the additive treatment. For infinitesimal excursions around a well-populated stationary state, freezing the diffusion coefficients at their steady-state values is consistent, and the small-signal calculation can remain accurate. In a direct time-domain simulation near or below threshold, the excursions are not small relative to the mean photon number and repeatedly sample the boundary at $n=0$. The neglected term $\delta{\bf x}\,d{\bf W}$ is then not perturbative, while the Gaussian approximation to the event counts also fails. Agreement between additive Langevin calculations and the discrete dynamics above threshold therefore does not extend continuously into the initial amplification regime. Because Eqs~\eqref{seqn1} and \ref{seqN2} follow from a jump process, their stochastic integrals have the It\^o interpretation. Rewriting them in the Stratonovich convention would require the corresponding noise-induced drift; changing conventions without that correction changes the model. Numerical implementations must also preserve the shared increment $dW_{\mathrm{em}}$ in the photon and carrier equations, with $dW_{\mathrm{em}}$ denoting the random emission into the lasing mode, including spontaneous and stimulated emission (the same increment appears with opposite signs in the photon and carrier equations because one emission event creates a photon and removes one excited carrier). These two requirements, together with non-negativity and adequate event counts per step, provide practical tests for any continuous approximation. The resulting physical picture is therefore scale-dependent. In the few-photon regime, the onset of lasing is a sequence of discrete, state-conditioned events, and the Laser Markov Chain is the appropriate description. As the photon population grows, many such events occur within each dynamical time scale, their Poisson statistics approach a Gaussian distribution, and the multiplicative coefficients vary only weakly about their stationary values. Additive Langevin noise is recovered as the leading large-population approximation, rather than as a separate physical source of fluctuations. Till now, the three coefficients $\eta_1$, $\eta_2$, and $\eta_3$ (or, equivalently, $\eta_a$, $\eta_b$, and $\eta_c$) are indeterminate. For the above analysis, we assume that after the lasing, the multiplicative noise tends to get closer and closer to the expressions determined by Henry, valid for macro-lasers \cite{Henry1983}. It is easily checked that the Henry expressions are recovered by setting:

\begin{align}\label{StCoeff1}
&\eta_1^2\rightarrow\frac{2D_{II}}{\tau_n<n^2>_{st.state}}=\frac{2{\mathcal R}I}{\tau_n<n^2>_{st.state}}\approx \frac{2{\mathcal R}}{\tau_nI}\nonumber\\
&\eta_2^2\rightarrow\frac{2(D_{II}+S)}{\tau_N<N^2>_{st.-state}}=\frac{2({\mathcal R}I+S)}{\tau_N<N^2>_{st.state}}\nonumber\\
&\approx \frac{2({\mathcal R}I+S)}{\tau_N<N>^{\!2}_{st.-state}}\nonumber\\
& \eta_3^2\rightarrow\frac{2D_{II}}{\sqrt{\tau_n\tau_N}<n N>_{st.state}}=\frac{2{\mathcal R}<n>_{st.state}}{\sqrt{\tau_n\tau_N}<nN>_{st.state}}\nonumber\\
&\approx\frac{2{\mathcal R}}{\sqrt{\tau_n\tau_N}<N>_{st.state}}
\end{align}
Here, $st.state$ stands for \textit{steady-state}. Moreover, ${\mathcal R}$, $I=<n>$, and $S$ denoting the \textit{Electron-hole recombination rate}, the \textit{mean intracavity photon number}, and the \textit{spontaneous emission rate}, respectively \cite{Henry1983}. According to Henry's theory, $D_{II}={\mathcal R}I$ is the \textit{photon-intensity variance-production rate}. Figure~\ref{figSNAdd} shows the stochastic evolution of the photon and carrier populations obtained with Henry's additive-noise model. In this case, noise produces only small fluctuations around the deterministic solution. The stochastic photon trajectory remains close to its deterministic counterpart, while the difference between the stochastic and deterministic carrier populations is barely visible. Figures~\ref{figSNMultPh} and~\ref{figSNMultCa} compare the photon and carrier dynamics obtained with multiplicative noise against the corresponding deterministic solutions. The difference is now much more pronounced. In particular, the median photon population departs substantially from the deterministic prediction. The carrier population also exhibits a clear, although smaller, deviation. Figure~\ref{figStSignals} displays the fluctuations of the stochastic photon signal generated by the QWS model with multiplicative noise. The blue shaded region represents the statistical spread of the photon signal, whereas the brown region indicates the intrinsic instrumental noise. All numerical results were obtained using $\Gamma_c=10^{9}sec^{-1}$, $\beta=0.1$, $\gamma=\gamma_c=10^9sec^{-1}$, $R=10^{12} electron\ sec^{-1}$, $\eta_1=10^9sec^{-1}$,  $\eta_2=10^9sec^{-1}$, $\eta_3=10^5sec^{-1}$, and $\chi=1.0$.

\section{Numerical Implementation of the Laser Markov Chain}\label{Comp}

\subsection{Stochastic Rate Equations}

The fully stochastic predictions for laser signal growth from noise are based on the recurrence relations originally introduced in~\cite{Puccioni2015} and subsequently refined:

\begin{eqnarray}
N_{q+1} & = & N_q + N_P - N_d \, , \\
S_{q+1} & = & S_q + E_S - L_S \, , \\
R_{q+1} & = & R_q + D_R - L_R \, , 
\end{eqnarray}
where $N$ denotes the carrier population, $S$ the stimulated photon number, and $R$ the spontaneous photon population coupled into the lasing mode. The terms $N_P$ represent pump carrier injection into the active region, $N_d$ accounts for spontaneous relaxation events, $E_S$ for stimulated emission events, while $L_S$ and $L_R$ denote cavity losses for stimulated and spontaneous photons, respectively (via the output mirror). Finally, $D_R$ represents spontaneous photons entering the lasing mode stochastically from the relaxation process $N_d$. The index $q$ specifies the population state at discrete time step $q$, with $q+1$ representing the subsequent step. All processes are inherently stochastic and implemented using Poissonian probabilities. To optimize computational efficiency, most Poissonian draws are approximated as Binomial distributions under the condition of small success probability (typically $\sim 10\%$), where the two distributions yield equivalent results.

\subsection{Simulation Parameters and Conditions}

\begin{table}[t]
\centering
{\caption{Stochastic processes and their numerical implementation using GLS libraries in C. $\mathbb P$ represents a Poissonian function call, while $\mathbb B$ stands for a binomial one.  The indices $i, j, k$ added to $\tau$ represent the actual times at which each event occurs, where the waiting time $\tau$ is reset to zero after a successful draw.}
\label{processes}}

\small
\setlength{\tabcolsep}{2.5pt}
\renewcommand{\arraystretch}{1.1}

\begin{tabular}{|c|c|c|c|c|}
\hline
Process & Event & R.V. & Distrib. & Probability \\
\hline\hline
Excit. 
& Pump Abs. 
& $N_p$ 
& $\mathbb{P}(P)$ 
& $P$ \\
\hline

Disexcit. 
& Pop. Relax. 
& $N_d$ 
& $\mathbb{B}(\mathcal{P}_N,N)$ 
& $1-e^{-\gamma\tau_i}$ \\
\hline

\multirow{4}{*}{Emiss.}
& Spont. (on-axis) 
& $D_L$ 
& $\mathbb{B}(\mathcal{P}_f,N_d)$ 
& $\beta$ \\
\cline{2-5}

& Stim. 
& $E_S$ 
& $\mathbb{B}(\mathcal{P}_{st},N)$ 
& $\gamma\beta(S+R)\Delta\tau$ \\
\cline{2-5}

& Spont. Loss 
& $L_R$ 
& $\mathbb{B}(\mathcal{P}_{T,L},R)$ 
& $1-e^{-\Gamma_c\tau_j}$ \\
\cline{2-5}

& Stim. Loss 
& $L_S$ 
& $\mathbb{B}(\mathcal{P}_{T,S},S)$ 
& $1-e^{-\Gamma_c\tau_k}$ \\
\hline
\end{tabular}
\end{table}

Simulations were executed with fixed cavity decay rate $\Gamma_c = 10^{11}\,\mathrm{s}^{-1}$ and carrier relaxation constant $\gamma = 10^9\,\mathrm{s}^{-1}$. Three distinct regimes were explored via the spontaneous emission factor $\beta$:

\begin{itemize}
\item For $\beta = 10^{-1}$ and $\beta = 10^{-3}$, each simulation runs for $100\,\mathrm{ns}$ with temporal sampling of $1\,\mathrm{ps}$, yielding $10^5$ data points per trajectory. Near threshold, pump events occur at frequencies around $1\%$.
\item For $\beta = 10^{-6}$, computational constraints necessitate shorter trajectories of $15\,\mathrm{ns}$ ($1.5 \times 10^3$ points) due to the significantly larger system populations required. In this regime, pump event frequencies reach approximately $10\%$, still within the range of validity of the substitution of a Binomial for a Poisson function.
\end{itemize}
In all cases, intracavity photon numbers are recorded at each sampling instant. Despite the higher event frequencies for small $\beta$, most probability evaluations return zero, preserving additivity and numerical reliability.

\subsection{Physical Scope and Validation}

\begin{figure}
  \centering
  \includegraphics[width=\linewidth]{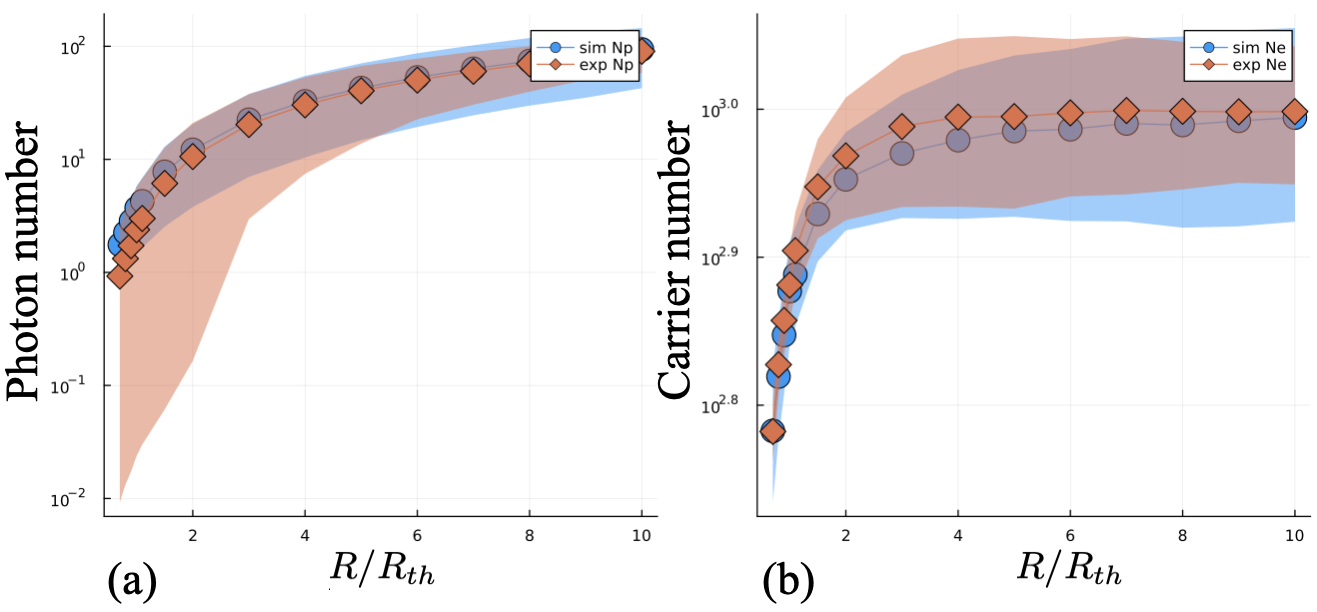} 
  \caption{Nanolaser:  $\beta = 10^{-1}$. Comparison of numerical simulations (brown) with predictions of the multiplicative stochastic model (blue) as functions of $R/R_{th}$. The labels “sim” (blue) and “exp” (brown) denote the model predictions and numerical results, respectively. Panels (a) and (b) show the photon number $n$ and carrier number $N$, respectively, with logarithmic vertical axes. Blue circles and brown diamonds indicate the mean values obtained from the model and numerical simulations, respectively. The corresponding standard deviations are represented by blue and brown shaded regions.}
  \label{figcomp}
\end{figure}

The numerical scheme implements Einstein's radiation-matter interaction picture for two-level systems~\cite{Einstein1917}. Recent theoretical work has demonstrated that this scheme constitutes a rigorous discretization of the Master Equation~\cite{Bundgaard2026}, analogous to probabilistic schemes developed in chemical physics~\cite{Gillespie1976,Gillespie1977}. No extrinsic noise is introduced; only intrinsic quantum stochasticity governs the dynamics. Non-radiative relaxation channels are omitted as they affect only the operating point relative to pump rate without altering fundamental energy exchanges. Experimental validation exists across multiple contexts: threshold identification~\cite{Wang2018}, pulse generation~\cite{Wang2019a}, and feedback dynamics~\cite{Wang2019b}. Superthermal photon statistics predicted by the model find correspondence at the mesoscale~\cite{Wang2020} with probable agreement also in nanolaser systems~\cite{Kreinberg2020}. Strict quantitative matching is not anticipated given that the model rests on general physical principles without device-specific construction details for commercial systems. Figure~\ref{figcomp} compares the numerical simulations with the predictions of the multiplicative stochastic model. The mean photon and carrier populations, together with their standard deviations, are shown as functions of the threshold-normalized pump rate $R/R_{\mathrm{th}}$. The model predictions (“sim”, blue) are compared with the LMC results (“exp”, brown), with the corresponding standard deviations represented by the shaded regions. Here, $N$ and $n$ denote the photon and electron numbers, respectively. Both quantities are displayed on logarithmic vertical scales. The two panels show the photon and electron populations for $\beta=10^{-1}$. The parameters used in the simulations are
\begin{equation}\label{comp1}
\Gamma_c\!=\!10^{11}\! sec^{-1},\ 
\gamma\!=\!\gamma_s\!=\!10^9\! sec^{-1},\ 
R_{\mathrm{th}}\!=\! 10^{12}\! sec^{-1}
\end{equation}
with
\begin{align}\label{comp2}
&\eta_1 = 7.5856\times 10^4sec^{-1},
\eta_2 = 5.902\times 10^3sec^{-1},\nonumber\\
&\eta_3 = 5.298sec^{-1}
\end{align}

\section{Discussion and conclusions}\label{concl}
Stimulated emission actually becomes statistically significant well before gain and loss balance out at the traditional laser threshold. The steady-state solutions make this clear, showing that $\beta$ heavily dictates the pump window where spontaneous and stimulated emission compete. In a macroscopic laser, this crossover happens within a tight pump range. For a high-$\beta$ nanolaser, however, the two processes coexist over a much wider interval. Yet, steady-state averages only tell part of the story. Crucially, tracking the mean photon population alone cannot tell whether the field is building up continuously or if it is driving an intermittent dynamic where spontaneous emission sporadically triggers cascades of stimulated events. Instead of guessing, the stochastic framework developed here through multiplicative noise builds directly from the elementary transitions driving the photon and carrier populations. Their rates map onto the system's instantaneous state: cavity loss scales with the internal photon count, while emission depends on the available carriers and, for stimulated emission, the number of photons already in the mode. Furthermore, an emission event simultaneously generates a photon and destroys an excited carrier. The resulting fluctuations are inherently state-dependent and anticorrelated. Applying a diffusion approximation to this jump process naturally yields multiplicative noise and a negative photon-carrier cross-correlation. To keep the associated diffusion matrix positive semidefinite, the strength of this correlation must obey a strict mathematical constraint. We use a numerical Laser Markov Chain as a discrete benchmark for our continuous stochastic framework. Instead of smoothing out fluctuations into continuous random forces, this approach samples individual events: pumping, carrier relaxation, spontaneous/stimulated emission, and cavity loss. This preserves the non-negative, integer nature of the populations while keeping the exact correlations born from individual transitions. Running simulations across a spectrum of $\beta$ values lets us track the stochastic dynamics all the way from the nanolaser regime up to macroscopic systems. Comparing the Laser Markov Chain against the multiplicative stochastic model shows that the continuous model successfully captures the statistics of both populations-provided the counts and event frequencies are high enough for Gaussian coarse-graining to make sense. Checking the mean populations and their fluctuations provides a direct numerical stress test for our state-dependent description. The excellent agreement proves that multiplicative noise is not just a phenomenological patch slapped onto deterministic rate equations. Instead, it is a mathematically rigorous continuous approximation of the underlying discrete transition statistics. This mapping also exposes exactly where the diffusion description breaks down. When the cavity holds only a handful of photons, a single emission or loss event alters the population by a fraction comparable to its total value. These steps are inherently discrete; they cannot accurately be mapped onto a continuous Gaussian process. Forcing Gaussian trajectories into this regime often causes them to cross the physical boundary at $n=0$. Relying on numerical fixes like clipping or reflecting negative values distorts the transition statistics and warps the physics. In this low-photon regime, the Laser Markov Chain, or the master equation it mirrors, is the only appropriate choice because it natively respects population boundaries and discrete events. A massive carrier reservoir alone cannot salvage a diffusion approximation; it needs a sufficiently high event count across every single active stochastic channel. At the other extreme, when the photon population climbs, a massive number of elementary events occur within a single dynamical timeframe, causing relative fluctuations to shrink. If these stochastic excursions stay small compared to the steady-state averages, the state-dependent diffusion matrix barely shifts over the system's trajectory. Evaluating this matrix strictly at the steady state yields a constant noise amplitude, returning directly to the familiar additive Gaussian forcing seen in linearized Langevin equations. Consequently, classic additive-noise models are best understood not as an independent noise source, but simply as the large-population, small-fluctuation limit of this same stochastic framework. Ultimately, these numerical results point to a unified hierarchy of stochastic descriptions. Right at the start of amplification, where isolated emission and loss events dictate the dynamics, a discrete Markov framework is mandatory. As populations and event counts grow, we can coarse-grain these discrete steps into a state-dependent, correlated diffusion process defined by multiplicative noise. Finally, well above threshold—where fluctuations are just minor ripples around a densely populated steady state- this multiplicative framework collapses back into the classic additive Langevin model. Choosing the right stochastic tool for laser amplification does not depend on an arbitrary "macroscopic" or "nanoscopic" label; it depends entirely on the populations, event rates, and relative scale of the fluctuations in the regime of interest.

\end{document}